\documentclass[preprint]{aastex}
\usepackage{amsmath,epsfig}
\newcommand{\beq}{\begin{equation}}
\newcommand{\eeq}{\end{equation}}
\newcommand{\bea}{\begin{eqnarray}}
\newcommand{\eea}{\end{eqnarray}}
\newcommand{\rem}[1]{ }

\begin{document}
\title{Dynamics of Astrophysical Bubbles and Bubble-Driven Shocks: Basic Theory, Analytical Solutions and Observational Signatures}

\author{Mikhail V. Medvedev\altaffilmark{1}} 
\affil{Department of Physics and Astronomy, University of Kansas, Lawrence, KS 66045}
\altaffiltext{1}{Also at the ITP, NRC ``Kurchatov Institute", Moscow 123182, Russia}

\author{Abraham Loeb} 
\affil{Astronomy Department, Harvard University, 60 Garden St., Cambridge, MA 02138}

\begin{abstract}
Bubbles in the interstellar medium are produced by astrophysical sources, which continuously or explosively deposit large amount of energy into the ambient medium. These expanding bubbles can drive shocks in front of them, which dynamics is markedly different from the widely used Sedov-von Neumann-Taylor blast wave solution. Here we present the theory of a bubble-driven shock and show how its properties and evolution are determined by the temporal history of the source energy output, generally referred to as the {\it source luminosity law}, $L(t)$. In particular, we find the analytical solutions for a driven shock in two cases: the self-similar scaling $L\propto (t/t_s)^p$ law (with $p$ and $t_s$ being constants) and the finite activity time case, $L\propto (1-t/t_s)^{-p}$. The latter with $p>0$ describes a  finite-time-singular behavior, which is relevant to a wide variety of systems with explosive-type energy release. For both luminosity laws, we derived the conditions needed for the driven shock to exist and predict the shock observational signatures. Our results can be relevant to stellar systems with strong winds, merging neutron star/magnetar/black hole systems, and massive stars evolving to supernovae explosions.
\end{abstract}

\section{Introduction}

Astrophysical bubbles are formed around sources with outflows in the form of winds, electromagnetic radiation, Poynting flux, etc., when outflows are strong enough to sweep up much material in the ambient medium. Examples include supernova remnants, pulsar wind nebulae, stellar-wind-driven bubbles around early type stars, Wolf-Rayet stars or star clusters \citep{GS06,C07, MM12}, as well as some models of gamma-ray bursts \citep{LB03}.  In a separate paper \citep{ML12}, we demonstrate that close binaries of compact objects (neutron stars, magnetars, black holes) can produce Poynting-flux-driven bubbles during their final inspiral and merger \citep{McW+L11,E+12}. The bubbles can be accompanied by shocks, which are known to be efficient accelerators of cosmic rays. Observationally, the shocks are detected via synchrotron radiation from the accelerated population of electrons in the ambient or self-generated magnetic fields, or by Compton up-scattering of lower-energy photons by the energetic electrons. Theoretical understanding the dynamics of the bubbles and the associated shocks combined with multi-wavelength telescope observations, gravitational wave search (e.g., with Advanced LIGO) and neutrino detection experiments (with KamiokaNDE, IceCube and others) can provide valuable information about the nature of the central engines.

In this paper we consider systems in which energy is continuously pumped into the medium by a central source. This energy produces excess pressure which pushes on the surrounding plasma and results in the formation of an expanding bubble (or a cavity). If the bubble expansion velocity exceeds the sound speed in the ambient medium, then a shock forms ahead of the bubble, as shown in Figure \ref{system}.  Such a shock wave is different from the Sedov-von Neumann-Taylor blast wave \citep{Sedov46} produced by a point-like instantaneous explosion and freely expanding into the interstellar medium (ISM). Instead, the shock is continuously driven by the ever-increasing pressure inside the bubble, rendering the classical Sedov-von Neumann-Taylor solution inapplicable. 

Formation and dynamics of astrophysical bubbles has been a subject of intense research (see e.g., the review by \citealp{OM88}). However, analytical studies have mostly been limited to the self-similar solutions $R(t)\propto t^\alpha$,\ $v(t)\propto t^{\alpha-1}$ with $R$ and $v$ being the size and the expansion speed, and $\alpha$  being a constant. This requires the energy deposition rate into the bubble, which we colloquially refer to as the ``source luminosity'' $dE_\text{bubble}/dt\equiv L(t)$, to be a power-law in time, $L(t)\propto t^p$, where $p$ is a constant. This is a very restricting assumption because many astrophysical sources exhibit an explosive behavior: they are nearly constant-luminosity sources for a long time followed by a rapid ---  ``explosive'' --- increase in energy output, formally tending to infinity within a finite time:\footnote{The singular (explosive) behavior occurs if $p>0$. However, our analysis below makes no assumption about the value of $p$.}
\beq
L(t)=L_s(1-t/t_s)^{-p},
\label{L-fts}
\eeq
where $t_s$ is the time during which the source is active, i.e., its ``lifetime'', $L_s$ is the source luminosity at early times $t\ll t_s$ and $p$ is the luminosity index, determining the rate of energy deposition in the bubble. Such a luminosity law with the finite-time singularity (FTS) is not rare in nature. For example, a binary consisting of two compacts objects, i.e., a neutron star (NS), a magnetar, or a black hole (least one of the companions should be magnetized) is producing a bubble filled with material with the relativistic equation of state \citep{ML12} and its luminosity is described by Eq. (\ref{L-fts}) with the index $p$ being around 3/2; the exact value depends on specific details of the energy extraction. The merger of two neutron stars is considered to be the conventional model of short gamma-ray bursts (GRBs). However, the formation of the bubble and the associated driven shock and their observational signatures have not been addressed. The bubble-driven shock can, however, produce a unique precursor to the main short GRB event and serve as a benchmark of the NS-NS progenitor. Combining these results with the detection of gravitational waves accompanied by a neutrino signal measured, respectively, by Advanced LIGO and other gravitational wave detectors (during the last moments of the binary inspiral) and by KamiokaNDE, IceCube and other neutrino detectors (during the merger event itself) will also greatly advance our understanding. Thus, the development of the theory of the bubble and shock dynamics in such systems is of great importance for astrophysics.

The paper is organized as follows. In Section \ref{s:model} we develop the theoretical model of a bubble and the bubble-driven shock and obtain the master equation describing the system evolution for an arbitrary luminosity law $L(t)$. In Section \ref{s:results} we find analytical solutions for the master equation for both the self-similar and singular (i.e., FTS) luminosity laws and explore the conditions under which the driven shock solutions exist. We also determined the conditions under which the exact FTS solutions (for the radius and velocity) exist. However, realistic systems, such as the neutron star  binaries, do not fall into that category. Nevertheless, we were able to find simple analytical approximate FTS solutions for them as well. Finally, we demonstrate that the obtained analytical solutions agree very well with the full numerical solutions in the appropriate limits. Finally, in Section \ref{s:concl} we present conclusions.

\section{The bubble+shock model}
\label{s:model}

\begin{figure}[b!]
\center
\includegraphics[angle = 0, width = 0.45\columnwidth]{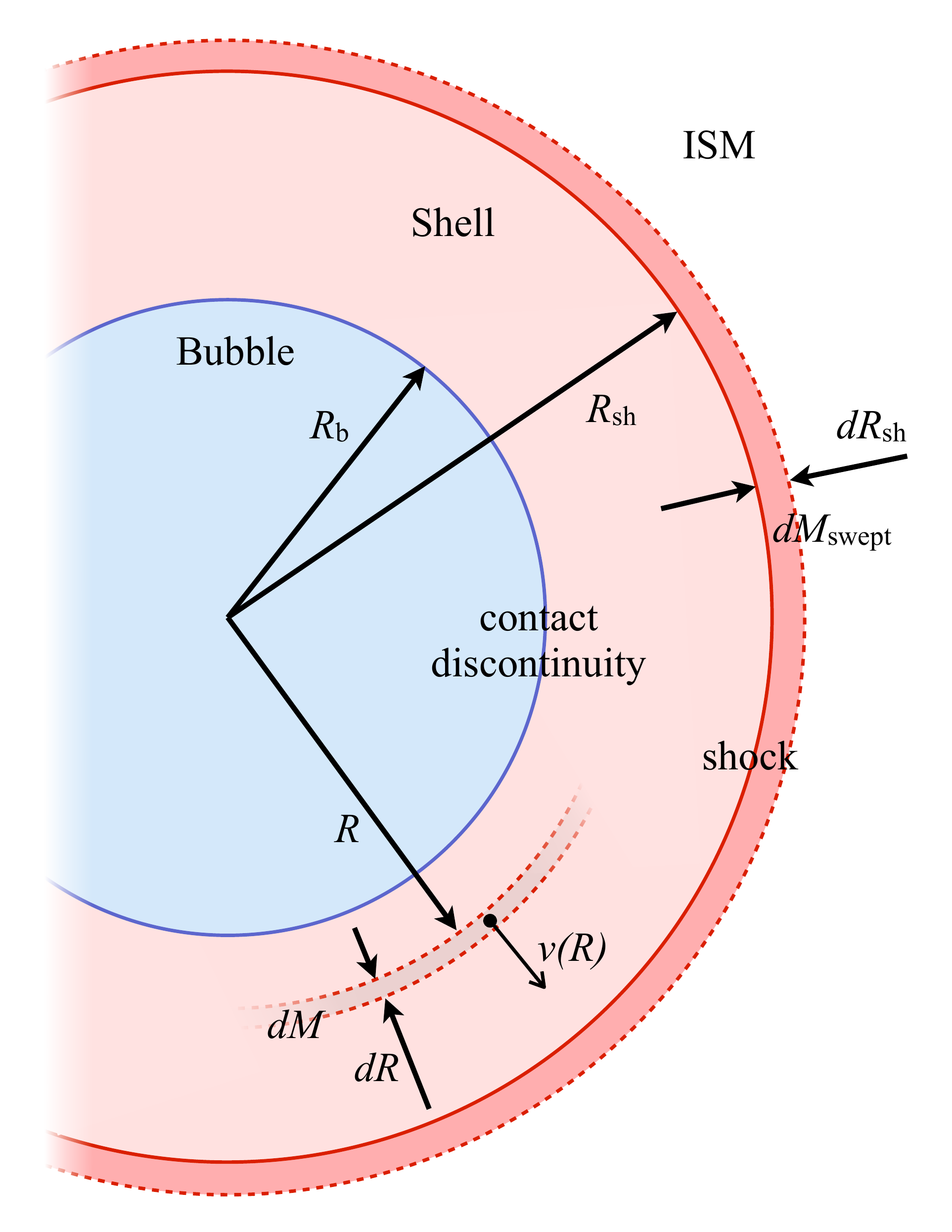}
\caption{Schematic representation of the system. The central source outflow produces a bubble, which creates a shock in the ambient medium and the shell of the shock-heated gas.}
\label{system}
\end{figure}

The system at hand is depicted in Fig. \ref{system}. For simplicity, we consider spherically symmetric systems. The bubble is filled with some material with pressure $p_\text{bubble}$ and the equation of state parameterized by the adiabatic index $\gamma_b$, which  depends on the bubble nature and composition. If the bubble is driven by Poynting flux (e.g., in a form of a strongly magnetized relativistic $e^-e^+$ wind) as in pulsar wind nebulae, or double neutron star or magnetar binaries, then the value of the index is $\gamma_b=4/3$. If the bubble is driven by a non-relativistic strong stellar wind, then $\gamma_b=5/3$.

The bubble surface acts as a piston and pushes the ambient plasma outward to produce an outgoing shock wave. Is the outgoing shock relativistic? Let us assume, for example, that the total energy $E\sim10^{46}~\text{erg}$ is deposited into the bubble over some time $t_\text{source}$ during which the source is active. If the shock is relativistic, it should expand with the speed $\sim c$ and the Lorentz factor
\beq
\Gamma\sim\frac{E}{Mc^2}\sim\frac{E}{(4\pi/3)(n_\text{ISM}m_p)(c t_\text{source})^3c^2}\gg1,
\eeq 
where $M$ is the mass of the ISM material swept by the shock over time $t_\text{source}$ and $m_p$ is the proton mass (assuming purely hydrogen ISM). For the ISM with a uniform density $n_\text{ISM}\sim1~\text{cm}^{-3}$, we have the constraint on the source lifetime:
\beq
t_\text{source}\ll4\times10^5E_{46}^{1/3}n_{\text{ISM},0}^{-1/3}~\text{s}\sim4.5\,E_{46}^{1/3}n_{\text{ISM},0}^{-1/3}~\text{days},
\eeq
where $E_{46}=E/(10^{46}~\text{erg})$, $n_{\text{ISM},0}=n_\text{ISM}/(1~\text{cm}^{-3})$ and similarly for other quantities henceforth. Thus, if the typical source activity is much longer than a few days, the shock can be assumed to be non-relativistic. This constraint is even less restrictive for lower-energy sources.

The shock is propagating into the unperturbed unmagnetized or weakly magnetized ISM with mass density $\rho_\text{ISM}$ and the adiabatic index $\gamma=5/3$. For simplicity, we assume the ISM to be cold, so that we can neglect its pressure; hence the shock is a high-Mach-number strong shock. Besides, weak shocks are less interesting from the observational point of view anyway: being just a mild perturbation to the medium, their observational signatures are hard to detect. Finally, between the shock and the bubble lies the shell of the shocked gas, which mass density and pressure, $\rho_\text{shell}$ and $p_\text{shell}$, are determined by the Rankine-Hugoniot shock jump conditions. The pressure equilibration time behind the shock is assumed to be fast enough to establish pressure equilibrium throughout the system, hence the bubble-shell interface plays a role of a contact discontinuity and $p_\text{bubble}=p_\text{shell}$.

We assume the central engine deposits energy (e.g., electromagnetic, kinetic, etc.) in the bubble with luminosity $L(t)\equiv dE/dt$, which is a function of time. This power goes into the internal energies (i) of the bubble, $dU_\text{bubble}$, and (ii) of the shocked gas shell, $dU_\text{shell}$, (iii) the change of the kinetic energy of the bulk motion of the shell, $dK_\text{shell}$, assuming its swept-up mass $M_\text{swept} =const$, (vi) the change of the kinetic energy, $dK_\text{@shock}$, of the newly swept gas, $dM_\text{swept}$, and also (v) heating of this shocked gas, $dU_\text{@shock}$, to satisfy the Rankine-Hugoniot relations. We can neglect the $p\,dV$ work due to the expansion because the external pressure is assumed to be vanishingly small in the cold ISM. Thus, the Master equation is 
\beq
L(t)\,dt=dU_\text{bubble} + dU_\text{shell} + \left.dK_\text{shell}\right|_{M_\text{swept}} 
+ dU_\text{@shock} + dK_\text{@shock}.
\label{master}
\eeq

In addition to this equation, the Rankine-Hugoniot relations for a strong shock should be used. The continuity equation yields
\beq
\frac{\rho_\text{shell}}{\rho_\text{ISM}}=\frac{u_1}{u_2}=\frac{v_\text{shock}}{v_\text{shock}-v_\text{shell}(R_\text{shock})}\simeq\frac{\gamma+1}{\gamma-1}\equiv \kappa
\eeq
and
\beq
\frac{v_\text{shell}(R_\text{shock})}{v_\text{shock}}=1-\kappa^{-1},
\eeq
where $\kappa$ is the constant compression ratio of a strong nonradiative shock, $u_1$ and $u_2$ are the upstream and post-shock velocities in the shock comoving frame which transform to the lab frame as $v_\text{shock}=u_1$ and $v_\text{shell}(R_\text{shock})=u_1-u_2$. From the momentum conservation, $p_\text{ISM}+\rho_\text{ISM}u_1^2=p_\text{shell}+\rho_\text{shell}u_2^2$, for a cold ISM (i.e., neglecting $p_\text{ISM}$), one gets
\beq
p_\text{shell}=\rho_\text{ISM}v_\text{shock}^2\left(1-\kappa^{-1}\right).
\label{pshell}
\eeq

We now calculate the terms in Eq. (\ref{master}).

\subsection{Internal energies}

The internal energies of the bubble and the post-shock gas shell are those of ideal gas:
\bea
dU_\text{bubble} &=& d\left(\frac{1}{\gamma_b-1}\ p_\text{bubble}V_\text{bubble}\right),\\
dU_\text{shell} &=& d\left(\frac{1}{\gamma-1}\ p_\text{shell}V_\text{shell}\right),
\eea
where $V_\text{bubble}=(4\pi/3) R_\text{bubble}^3$ and $V_\text{shell}=V_\text{shock}-V_\text{bubble}$ are the volumes of the bubble and the shell, respectively, and $V_\text{shock}=(4\pi/3) R_\text{shock}^3$ and recall that $p_\text{bubble}=p_\text{shell}$. The swept-up ISM mass occupies the post-shock shell, $M_\text{swept}=\rho_\text{ISM}V_\text{shock}=\rho_\text{shell}V_\text{shell}$, hence one has the relations:
\beq
\frac{V_\text{shell}}{V_\text{shock}}=\kappa^{-1}, \qquad \frac{V_\text{bubble}}{V_\text{shock}}=1-\kappa^{-1}.
\eeq
Therefore,
\bea
dU_\text{bubble} &=& \frac{\left(1-\kappa^{-1}\right)^2}{\gamma_b-1}\ \rho_\text{ISM}\ d\left(v_\text{shock}^2 V_\text{shock}\right),
\label{dUbubble}\\
dU_\text{shell} &=& \frac{\left(1-\kappa^{-1}\right)\kappa^{-1}}{\gamma-1}\ \rho_\text{ISM}\ d\left(v_\text{shock}^2 V_\text{shock}\right).
\label{dUshell}
\eea

\subsection{Kinetic energy}

To calculate the shell kinetic energy one needs to know the post-shock velocity profile $v_\text{shell}(R)$. The post-shock gas is hot, hence thermal conduction is rapid and the temperature can be taken uniform throughout the shell. Since both temperature and pressure are uniform, then density is uniform as well, $\rho(R)=\rho_\text{shell}=\rho_\text{ISM}\kappa=const$. Continuity equation of an incompressible gas in a steady state, $\nabla\cdot{\bf v}=\partial_R\left(R^2 v_\text{shell}(R)\right)=0$, together with the boundary condition $v(R_\text{shock})=v_\text{shell}(R_\text{shock})=v_\text{shock}\left(1-\kappa^{-1}\right)$ yields
\beq
v_\text{shell}(R)=v_\text{shell}(R_\text{shock})\left(\frac{R_\text{shock}}{R}\right)^2.
\eeq

The kinetic energy of the shocked gas (see Fig. \ref{system}) is
\beq
K_\text{shell}=\int_{0}^{M_\text{swept}}\frac{dM v_\text{shell}(R)^2}{2}=\frac{M_\text{swept}v_\text{shock}^2}{2}\,\xi,
\eeq
where 
\bea
\xi &=& 3r\left(1-\kappa^{-1}\right)^2\left[\left(1-\kappa^{-1}\right)^{-1/3}-1\right]
\nonumber\\
&\simeq& \frac{12}{\gamma^2-1}\left[\left(\frac{\gamma+1}{2}\right)^{1/3}-1\right]
=\frac{9}{4}\left(6^{2/3}-3\right)
\eea
and we have used that $dM=(\rho_\text{ISM}\kappa)\, 4\pi R^2dR$ and $R_\text{bubble}/R_\text{shock}=\left(V_\text{bubble}/V_\text{shock}\right)^{1/3}=\left(1-\kappa^{-1}\right)^{1/3}$. At last, the change of the shell kinetic energy due to acceleration/deceleration is
\beq
\left.dK_\text{shell}\right|_{M_\text{swept}} = \rho_\text{ISM}(4\pi/3) R_\text{shock}^3\,\xi\ d\!\left(v_\text{shock}^2/2\right).
\label{dKshell}
\eeq

\subsection{Shock contributions}

As the shock propagates, it heats up the cold ISM gas of mass $dM_\text{swept}=\rho_\text{ISM}\, dV_\text{shock}$ to the post-shock temperature $T_\text{shell}=p_\text{shell}/n_\text{shell}=m_p v_\text{shock}^2\left(1-\kappa^{-1}\right)\kappa^{-1}$ and also accelerates it to the post-shock velocity, $v_\text{shell}(R_\text{shock})=v_\text{shock}\left(1-\kappa^{-1}\right)$. Thus, at the shock
\bea
dU_\text{@shock} &=& \frac{\left(1-\kappa^{-1}\right)\kappa^{-1}}{\gamma-1}\,\rho_\text{ISM}\,dV_\text{shock} v_\text{shock}^2,
\label{dUshock}\\
dK_\text{@shock} &=& \rho_\text{ISM}\,dV_\text{shock} \frac{v_\text{shock}^2}{2}\left(1-\kappa^{-1}\right)^2.
\label{dKshock}
\eea

\subsection{Final analysis}

Equation (\ref{master}) together with equations (\ref{dUbubble}), (\ref{dUshell}), (\ref{dKshell}), (\ref{dUshock}) and (\ref{dKshock}) determine the evolution of a shock driven by the pressure inside a cavity. Given the luminosity, $L(t)$, it allows us to determine $R_\text{shock}(t)$, because the shock velocity is $v_\text{shock}=dR_\text{shock}/dt\equiv\dot R_\text{shock}$ and the shocked volume is $V_\text{shock}=(4\pi/3) R_\text{shock}^3$. All other quantities, e.g., $R_\text{bubble},\ p_\text{shell}$, etc. follow straightforwardly from the equations above. Hereafter we will often omit the subscript ``shock'' wherever it does not cause a confusion. Upon the substitutions, Eq. (\ref{master}) becomes
\beq
\frac{L(t)}{(4/3)\pi\rho_\text{ISM}} =
\eta\frac{d(R^3\dot R^2)}{dt}+\frac{\xi}{2}R^3\frac{d\dot R^2}{dt}+\zeta \dot R^2\frac{d R^3}{dt}
\label{master2}
\eeq
where
\bea
\eta &=& \frac{\left(1-\kappa^{-1}\right)^2}{\gamma_b-1}+\frac{\left(1-\kappa^{-1}\right)\kappa^{-1}}{\gamma-1}
\nonumber\\
&\simeq& \frac{2(\gamma_b+1)}{(\gamma+1)^2(\gamma_b-1)}=\frac{63}{32}, 
\\
\zeta &=& 3\left[\frac{\left(1-\kappa^{-1}\right)\kappa^{-1}}{\gamma-1}+\frac{\left(1-\kappa^{-1}\right)^2}{2}\right]
\nonumber\\
&\simeq&\frac{12}{(\gamma+1)^2}=\frac{27}{16}.
\eea
Here the first term represents the internal energies of the bubble and shell, the second term is the bulk kinetic energy of the shell and the last term is the contribution of the shock. Hereafter we use, for concreteness, that the bubble is filled with relativistic material (e.g., a relativistic plasma, or a highly magnetized wind, or electromagnetic radiation) with $\gamma_b=4/3$, the ambient gas is non-relativistic with $\gamma=5/3$ and the compression ratio is $\kappa\simeq4$.

Master equation (\ref{master2}) is the inhomogeneous second-order nonlinear differential equation, which can further be simplified to yield
\beq
\frac{L(t)}{(4\pi/3)\rho_\text{ISM}} =(3\eta+\zeta)R^2\dot R^3+(2\eta+\xi)R^3\dot R \ddot R.
\label{master3}
\eeq
This is the main equation of our analysis. For any function of the source emission luminosity, $L(t)$, pumping energy into the bubble, this equation describes the evolution of the shock radius, $R_\text{shock}\equiv R(t)$, and the associated parameters, e.g., the shock velocity, $v(t)=\dot R(t)$, the size of the bubble, $R_\text{bubble}(t)=\left(1-\kappa^{-1}\right)^{1/3}R(t)\simeq(3/4)^{1/3}R(t)$, etc. We stress that this equation is applicable to a bubble of {\em any} origin. For example, it describes a Poynting-flux-driven bubble formed by an inspiraling binary; in this case $\gamma_b=4/3$. It also describes a bubble blown by a strong stellar wind with the kinetic energy luminosity given by $L(t)$ and the adiabatic index of the gas in the bubble $\gamma_b=5/3$.

\section{Results and observational predictions}
\label{s:results}

To proceed further, one needs to specify the {\em source luminosity law}, $L(t)$. Below, we will consider the two most common ones: the self-similar law $L(t)\propto t^p$ and the ``explosive'' finite-time-singular law $L(t)\propto (t_s-t)^{-p}$ with $p>0$,

\subsection{Self-similar solution}
\label{s:sss}

\subsubsection{Structure}

We first look for a self-similar solution. Let us assume that the source luminosity and the shock position are power-law functions of time
\beq
L(t)=L_s (t/t_s)^p,\qquad R(t)=R_s (t/t_s)^\alpha,
\label{selfsim}
\eeq
where $R_s$ and $\alpha$ are constants to be determined and $L_s,\ t_s$ and $p$ are known constants set by the source physics. This self-similar solution is valid for the duration of the source activity, i.e., while the bubble pressure is high enough to push the shock; at (much) later times, the shock dynamics should asymptote the Sedov-von Neumann-Taylor solution. 

The self-similar solution (\ref{selfsim}) to equation (\ref{master3}) is
\beq
\alpha=\frac{p+3}{5}, \qquad R_s=\left(\frac{3}{4\pi}\frac{L_st_s^3}{\rho_\text{ISM}A}\right)^{1/5},
\label{ss-Rs}
\eeq
where 
\bea
A &=& (3\eta+\zeta)\alpha^3+(2\eta+\xi)\alpha^2(\alpha-1)
\nonumber\\
&\simeq& \frac{9}{4}\alpha^2\left[\frac{5}{4}-6^{2/3}+\alpha\left(\frac{17}{8}+6^{2/3}\right)\right].
\label{A}
\eea
The scalings given in equations (\ref{selfsim}), (\ref{ss-Rs}) agree with those in \citep{OM88}. The above solution is meaningful if $R>0$, i.e., $A>0$ and, hence,
\beq
\alpha>\alpha_\text{crit}=\frac{6^{2/3}-5/4}{6^{2/3}+17/8}\simeq0.378
\eeq
and
\beq
p>p_\text{crit}=5\alpha_\text{crit}-3\simeq-1.11.
\eeq
This condition means that the energy injection in the system cannot be too slow; otherwise the shock would move too fast for the contact discontinuity to catch up with it and the assumption of the pressure equilibration breaks down.

\subsubsection{Estimates}

We have obtained that given the luminosity of the source, $L(t)=L_s(t/t_s)^p$, the shock evolution is given by $R_\text{shock}(t)=R_s(t/t_s)^\alpha$ and $v_\text{shock}(t)=(\alpha R_s/t_s) (t/t_s)^{\alpha-1}$ with $\alpha=(p+3)/5$. The value of $R_s$ is rather insensitive to the numerical value of $A$ (unless $\alpha$ is very close to the critical value), so, $R_s\sim(L_s t_s^3/\rho_\text{ISM})^{1/5}$ is a rather good order-of-magnitude estimate. More accurately, assuming the source activity to be of the order of a hundred days, $t_s\sim10^{7}$~s, and the luminosity $L_s\sim10^{39}~\text{erg s}^{-1}$, we estimate that 
\bea
R_\text{shock}(t_s) &=& R_s\simeq4.3\times10^{16}\ A^{-1/5}L_{s,39}^{1/5}t_{s,7}^{3/5}n_{\text{ISM},0}^{-1/5}~\text{cm},
\label{ss-Rshock}\\
v_\text{shock}(t_s) &=& \alpha R_s/t_s\simeq4.3\times10^{9}\ \alpha A^{-1/5} L_{s,39}^{1/5}t_{s,7}^{-2/5}n_{\text{ISM},0}^{-1/5}~\text{cm s}^{-1}.
\eea
If $\alpha<1$, then the shock velocity $\propto t^{\alpha-1}$ is decreasing with time, thus at some earlier time it was $\sim c$, so the non-relativistic approximation was invalid. For the solution to be valid at early times for the duration of the source activity, the shock velocity should be increasing with time, i.e., $\alpha$ should be greater than unity, hence the luminosity index must be $p>2$.

\subsubsection{Observable signature}
\label{s:obs}

We assume that the shock accelerates electrons via Fermi process and generates/amplifies magnetic field. The relativistic electrons in magnetic fields produce synchrotron radiation which can be observed. We assume that the electrons and magnetic fields carry, respectively, fractions $\epsilon_e$ and $\epsilon_B$ of the internal energy density of the shocked gas, cf., Eq. (\ref{dUshell}), 
\beq
u_\text{shell}=\frac{1-\kappa^{-1}}{\gamma-1}\ \rho_\text{ISM}v_\text{shock}^2,
\eeq
that is
\beq
\bar\gamma_em_ec^2n_{e,\text{shell}}=\epsilon_e u_\text{shell}, \qquad
{B^2}/{8\pi}=\epsilon_B u_\text{shell}, 
\eeq
where $\bar\gamma_em_ec^2$ is the average energy of an electron and $n_{e,\text{shell}}=\kappa\,n_{e,\text{ISM}}\simeq \kappa\,n_\text{ISM}$ is the number density of electrons in the shocked gas shell. Numerically, the average Lorentz factor of accelerated electrons is 
\bea
\bar\gamma_e(t) &=& \epsilon_e\frac{\kappa-1}{\kappa^2(\gamma-1)}\frac{m_p}{m_e}\left(\frac{v_\text{shock}}{c}\right)^2
\nonumber\\
&\simeq&11\ \epsilon_e\alpha^2 A^{-2/5} L_{s,39}^{2/5}t_{s,7}^{-4/5}n_{\text{ISM},0}^{-2/5}t_{7}^{2(\alpha-1)},
\eea
so the bulk electrons are mildly relativistic $\bar\gamma_e\sim3$ for a typical acceleration efficiency of $\epsilon_e=0.3$. If the radiating electrons are distributed in energy as a power-law with index $s$ as $d n_e/d\gamma\propto \gamma^{-s}$ with a minimum Lorentz factor $\gamma_m$, then $\gamma_m=\bar\gamma_e(s-2)/(s-1)$. The magnetic field strength is
\bea
B(t) &=& \left(\epsilon_B\frac{\kappa-1}{\kappa(\gamma-1)}8\pi m_p n_\text{ISM} v_\text{shock}^2\right)^{1/2}
\nonumber\\
&\simeq& 
3.0\times10^{-2}\ \epsilon_B^{1/2} \alpha A^{-1/5} L_{s,39}^{1/5}t_{s,7}^{-2/5}n_{\text{ISM},0}^{3/10}t_{7}^{\alpha-1}~\text{G},
\eea 
that is, $B$ is of the order of a milligauss for a nominal $\epsilon_B\sim10^{-3}$, which is much larger than the typical microgauss fields in the ISM. This means that the field should be generated or amplified at the shock by an instability (Weibel, Bell, cyclotron, and such), or by MHD turbulence, or by some other mechanism. 

Relativistic electrons emit synchrotron radiation. The peak frequency of the radiation spectrum is in the radio band,
\bea
\nu_s(t) &=& (2\pi)^{-1}\gamma_m^2\,\frac{eB}{m_e c}
\nonumber\\
&\simeq& 9.3\times10^6\ \epsilon_e^2\epsilon_B^{1/2} \alpha^5 A^{-1} [(s-2)/(s-1)]^2 L_{s,39}t_{s,7}^{-2}n_{\text{ISM},0}^{-1/2}t_{7}^{5(\alpha-1)}
~\text{Hz}.
\eea
The total emitted power by a relativistic electron is $P=(4/3)\sigma_T c \bar\gamma_e^2(B^2/8\pi)$, where $\sigma_T$ is the Thompson cross-section. Thus the spectral power at the peak (measured in erg~s$^{-1}$~Hz$^{-1}$) is $P_{\nu,\text{max}}(t)\approx P/\nu_s\simeq(\sigma_T m_e c^2 /3e)B$. The observed spectral peak flux from a source located in our galaxy at the distance $D=10^{23}~\text{cm}$ (i.e., $\sim 30$~kpc) is $F_{\nu,\text{max}}=N_e P_{\nu,\text{max}}/(4\pi D^2)$, where $N_e=n_{e,\text{ISM}}V_\text{shock}$ is the total number of emitting electrons, hence
\bea
F_{\nu,\text{max}}(t) 
&=& \frac{1}{4\pi D^2}\left(\frac{4\pi}{3} R_\text{shock}^3 n_\text{ISM}\right)\frac{\sigma_T m_e c^2}{3e}B
\nonumber\\
&\simeq& 3.0\times10^{3}\ D_{23}^{-2}\epsilon_B^{1/2} \alpha A^{-4/5} L_{s,39}^{4/5}t_{s,7}^{7/5}n_{\text{ISM},0}^{7/10}t_{7}^{4\alpha-1}~\text{Jy}.
\eea
Although the peak frequency can fall in the self-absorbed part of the spectrum, as is typical of supernova shocks (e.g., \citealp{WL99}), the peak flux above is still useful for the normalization of the spectrum  
\bea
F_\nu(\nu,t) &=& F_{\nu,\text{max}}(t)\,\left(\frac{\nu}{\nu_s(t)}\right)^{-(s-1)/2}
\nonumber\\
&\propto&\nu^{-(s-1)/2}t^{(3-5s+\alpha(3+5s))/2}
\nonumber\\
&\propto&\nu^{-0.75}t^{7.75\alpha-4.75},
\eea
the latter scaling corresponds to the nominal value of $s=2.5$. We remind that the index $\alpha$ is related to that of the energy injection $L(t)\propto t^p$ as $\alpha=(p+3)/2$.  

Let us write $F_\nu(\nu,t)\propto\nu^a t^b$, where $a$ and $b$ are spectral and temporal indexes determined from observations. Then one can infer the (effective) energy injection index as
\beq
p=(1-5a+b)/(4-5a).
\eeq
Also, for a non-self-similar luminosity law, one can define the effective index as
\beq
p_\text{eff}=d\log{L(t)}/d\log t,
\eeq
which can be compared with observations. 

Finally, we estimate the self-absorption frequency as the frequency at which the optical depth of the emitting region of thickness $R_\text{shock}-R_\text{bubble}$ is of the order of unity, $\tau_\nu=\kappa_\nu(R_\text{shock}-R_\text{bubble})\sim1$, where $\kappa_\nu$ is the self-absorption coefficient \citep{RL}:
\beq
\kappa_\nu=-\frac{1}{8\pi m_e\nu^2}\int d\gamma P_\nu(\gamma)\gamma^2\frac{\partial}{\partial\gamma}\left(\frac{1}{\gamma^2}\frac{dn_e}{d\gamma}\right).
\eeq
A power-law distributed electrons $dn_e/d\gamma=(n_{e,\text{shell}}/\gamma_m)(\gamma/\gamma_m)^{-s}$ produce the spectrum (per electron) $P_\nu(\gamma)\simeq P_{\nu,\text{max}}f(\nu/\nu_c)$, where $f(x)$ is the function $F(x)$ of \citet{RL} involving the integral of a modified Bessel function, up to a numerical factor. Using a dimensional analysis, noting that $n_{e,\text{shell}}(R_\text{shock}-R_\text{bubble})\simeq n_\text{ISM}R_\text{shock}/3$ and absorbing all numerical factors into a constant $K\sim10^{-2}$, one gets the absorption frequency, $\nu_a$, 
\beq
\nu_a\simeq\left(K\sigma_Tc\bar\gamma_en_\text{ISM}R_\text{shock}/m_e\right)^{2/(s+4)}\nu_s^{(s-2)/(s+4)}.
\eeq
which depends very weakly of the constant $K$, the density and the shock radius and is nearly independent of the synchrotron frequency for typical spectral indices $s\gtrsim2$. For typical values of the parameters, $\nu_a$ is of the order of $\sim10^8-10^9~\text{Hz}$. Note that this analysis assumes that all the electrons are relativistic; otherwise, if only a fraction $\eta_\text{rel}<1$ of them is relativistic, the self-absorption frequency is lower by $\sim\eta_\text{rel}^{2/(s+4)}$.

\subsection{Solution with a finite-time singularity}
\label{s:ftss}

\subsubsection{General structure}

Traditionally in astrophysics, one looks for a self-similar solution, which is presented above. However, for some astrophysical systems, self-similarity may not be a good approximation. For example, in the paper \citep{ML12} we have shown that the evolution of the binary system is described by a finite-time singular (FTS) solution, because the inspiral takes a finite time. We have found that the luminosity is
\beq
L(t)=L_s (1-t/t_s)^{-p_*},
\label{finitetime}
\eeq
where $L_s,\ t_s$ and $p_*$ are known ``source" constants set by its physics and initial conditions. To be precise, these scaling should break down at some earlier time $t_m<t_s$, otherwise the total energy diverges if $p_*>1$, which is the case of a neutron star binary, for example. It also breaks down on physics grounds because general relativistic and other effects, as well as finite object sizes, were omitted from consideration. The above form of $L(t)$ suggests to look for a solution in the form
\beq
R(t)=R_{s,*} (1-t/t_s)^{-\alpha_*},
\label{fts-Rs}
\eeq
where $R_s$ and $\alpha_*$ are constants to be determined. Such a solution is formally valid at times $t<t_s$. At late times $t\gg t_s$, there is no energy injection, therefore the shock dynamics should asymptote the Sedov-von Neumann-Taylor solution. 

The finite-time singular solution, Eq. (\ref{fts-Rs}), to the Master equation (\ref{master3}) is
\beq
\alpha_*=\frac{p_*-3}{5}, \qquad R_{s,*}=\left(\frac{3}{4\pi}\frac{L_st_s^3}{\rho_\text{ISM}A_*}\right)^{1/5},
\label{soln}
\eeq
where 
\bea
A_* &=& (3\eta+\zeta)\alpha_*^3+(2\eta+\xi)\alpha_*^2(\alpha_*+1)
\nonumber\\
&\simeq& \frac{9}{4}\alpha_*^2\left[-\frac{5}{4}+6^{2/3}+\alpha_*\left(\frac{17}{8}+6^{2/3}\right)\right].
\eea
The solution is physically meaningful if $R>0$, hence $A_*>0$,
\beq
\alpha_*>\alpha_{*,\text{crit}}=\frac{5/4-6^{2/3}}{17/8+6^{2/3}}\simeq-0.378
\eeq
and
\beq
p_*>p_{*,\text{crit}}=5\alpha_{*,\text{crit}}+3\simeq-1.11.
\eeq
This condition constrains the energy injection index: if $p_*<p_{*,\text{crit}}$ then the energy injection rate is not enough for the contact discontinuity to catch up with the shock. 

Unlike the self-similar solution, this solution with a finite time singularity has another constraint. The physically meaningful solution (in this particular class of solutions) is the one of the {\em expanding} shock; hence
\beq
\alpha_*>0 ~\text{ and }~ p_*>3.
\eeq
We emphasize that the above scalings are applicable for the duration of the central engine activity only, $t< t_s$. Moreover, the solution discussed in the previous subsection does not have much physical meaning, because the system size increases to infinity in a finite time. Perhaps, it can make sense at times substantially smaller than $t_s$, when relativistic and other effects are negligible.  

As we mentioned earlier, some astrophysical systems, such as the neutron star and magnetar binaries \citep{ML12}, have the luminosity indexes $p_*=3/2$ or 7/4, depending on the electromagnetic energy extraction mechanism. For such $p_*$ values from in the range, $p_{*,\text{crit}}<p_*<3$, the bubble-driven shock exists, but the solution given by equation (\ref{fts-Rs}) describes an unphysical collapsing shock with $R(t)\to0$ as $t\to t_s$. This result is independent of most of the assumptions and easily follows from the energetics considerations or just the dimensional analysis, cf. equation (\ref{master3}),
\beq
(1-t/t_s)^{-p_*}\propto L(t)\sim d(Mv^2)/dt\sim \rho\, d(R^3\dot R^2)/dt\propto (1-t/t_s)^{-5\alpha_*-3},
\eeq 
which yields the relation $p_*=5\alpha_*+3$. On physics grounds, for values $p_{*,\text{crit}}<p_*<3$, the shock radius tends to a constant, $R\to R_\text{max}$ as $t\to t_s$, but the form $(1-t/t_s)^{-\alpha_*}$ does not have such asymptotic for any $\alpha_*\not=0$. A different form is needed, but it seems unlikely that one can find a general or partial solutions to Eq. (\ref{master3}) directly, because the solution should be of the form $\int dt\sqrt{(1-t)^{-p_*+1}+C}$, which cannot be expressed in elementary functions for an arbitrary $p_*$, though it can be expressed in elliptic functions for some rational $p_*$. Below we obtain an approximate solution for such a regime.

\subsubsection{Approximate realistic FTS solution}
\label{s:ars}

In the previous section, we have found that the bubble-driven shock can exist in systems with the luminosity index $p_*>p_{*,\text{crit}}$, but if $p_*$ is in the range $p_{*,\text{crit}}<p_*<3$, the physically meaningful solutions are not described by the pure FTS solutions. Instead, the physical FTS solution should have the property that $R_\text{shock}\to R_\text{max}=const$, but $v_\text{shock}\propto (1-t/t_s)^{-\alpha^*-1}\to\infty$ as $t\to t_s$, with $\alpha^*$ being a new constant. The general and/or exact analytical solution of this kind does not exist, so we construct here a composite approximate solution.

First, we notice that the dependence $L(t)=L_s (1-t/t_s)^{-p_*}$ implies a constant luminosity source at early times, $t\ll t_s$, that is $L(t)\simeq L_s (t/t_s)^0$. This case corresponds to the self-similar solution with $p=0$ in Section \ref{s:sss} for which the solution exists:
\beq
R_\text{shock}(t)=R_s(t/t_s)^{3/5}\to R_s, ~\text{ as }~ t\to t_s,
\label{Rsh}
\eeq
where $R_s$ is given by equations (\ref{ss-Rs}) or (\ref{Rshock}). Second, we make substitutions 
\beq
R=R_s, \quad \dot R\equiv v_\text{shock}(t)= v_s (1-t/t_s)^{-\alpha^*-1},
\eeq
where $v_s$ is a constant to be determined. By examining of the right hand side of equation (\ref{master3}), one sees that both terms are divergent but the last term dominates as $t\to t_s$, because $p_*<3$ and hence $\alpha^*<0$. Keeping the leading term, one has
\beq
\alpha^*=\frac{p_*-3}{2}, \qquad v_s=\left(\frac{3}{4\pi}\frac{L_st_s}{\rho_\text{ISM}(2\eta+\xi) R_s^3}\right)^{1/2}
=\left(\frac{A}{2\eta+\xi}\right)^{1/2}\frac{R_s}{t_s},
\label{FTSfactors}
\eeq 
where $A\simeq4.3$ is given by equation (\ref{A}) with $\alpha=3/5$ and the prefactor $\left[{A}/{(2\eta+\xi)}\right]^{1/2}\simeq0.97$. 

Thus, we have found an approximate solution describing evolution of the bubble-driven shock at the late times:
\bea
v_\text{shock}(\Delta t)&\simeq&v_s (\Delta t/t_s)^{-(p_*-1)/2}, 
\label{vs}\\
R_\text{shock}(\Delta t)&\simeq&R_s(\Delta t/t_s)^{0}=const,
\label{Rs}
\eea
where we introduced a new variable $\Delta t=(t_s-t)\to0$, which is more convenient when $t\lesssim t_s$. Note that the obtained regime is rather interesting: the shock is rapidly accelerating but its size and swept-up mass remain nearly constant.

\subsubsection{Estimates}

The characteristic values of the shock radius and velocity in the solution above are given by equations (\ref{vs}) and (\ref{Rs}). However, these values follow from an approximate analytical analysis. Comparison with the exact numerical solutions (Section \ref{s:ns}) yields {\it ad hoc} correction factors for $R_s$ and $v_s$, see Eqs. (\ref{chiR}) and (\ref{chiV}), namely $\chi_R\simeq1.33,\ \chi_v\simeq0.51$. We use these values in the estimates below. We have
\bea
R_s &\simeq& 3.2\times10^{16}\ \chi_R\, L_{s,39}^{1/5}t_{s,7}^{3/5}n_{\text{ISM},0}^{-1/5}~\text{cm},
\label{Rshock}\\
v_s &\simeq& 3.1\times10^{9}\ \chi_v\, L_{s,39}^{1/5}t_{s,7}^{-2/5}n_{\text{ISM},0}^{-1/5}~\text{cm s}^{-1},
\eea
where we assumed $\gamma_b=4/3$ and $\gamma=5/3$, for concreteness. We, thus, have 
\bea
v_\text{shock}(\Delta t) 
&\simeq&3.1\times10^{8+p_*}\ \chi_v\, L_{s,39}^{1/5}t_{s,7}^{-(9-5p_*)/10}n_{\text{ISM},0}^{-1/5}\Delta t_5^{-(p_*-1)/2}~\text{cm s}^{-1}
\nonumber\\
&\simeq&9.8\times10^{9}\ \chi_v\, L_{s,39}^{1/5}t_{s,7}^{-3/20}n_{\text{ISM},0}^{-1/5}\Delta t_5^{-1/4}~\text{cm s}^{-1},
\eea
where we used a nominal value $p_*\simeq3/2$. Note that if  $p_*>1$ then $v_\text{shock} \propto (\Delta t/t_s)^{-(p_*-1)/2}\to \infty$, as $\Delta t\to 0$ and $v_\text{shock}$ approaches the speed of light at times 
\beq
\Delta t\lesssim 10^{7-2/(p_*-1)}\ t_{s,7}\left(\chi_v\, L_{s,39}^{1/5}t_{s,7}^{-2/5}n_{\text{ISM},0}^{-1/5}\right)^{2/(p_*-1)}~\text{s},
\eeq
that is, about $10^3$~s before the ``explosion" time $t_s$, for $p_*=3/2$. At these times our assumption of the nonrelativistic shock breaks down and a different analysis is needed. Note also that at this time, the dynamical time of the bubble needed to establish pressure equilibrium throughout is longer, $R_s/c\sim 10^6$~s, so an accurate analysis should instead be involved to find a full dynamical solution. Such a consideration goes beyond the scope of the present paper. 

Regardless of the model assumption used, plasma processes and details of particle acceleration impose additional constraints as follows. If the characteristic dynamical time of the system, which is $\sim\Delta t$, is longer than the inverse collisional frequency $\nu_\text{coll}^{-1}$ then a collisional shock forms. Otherwise,  when $\Delta t<\nu_\text{coll}^{-1}$ then the shock is collisionless. In this case, if $\Delta t$ is (much) shorter then the Larmor frequency in the ambient field, the shock structure is not sensitive to the ISM field, but, instead, is determined by kinetic plasma instabilities (e.g., electrostatic Buneman or electromagnetic Weibel-like ones driven by particle anisotropy at the shock). The shortest associated timescale is the ion plasma time $\omega_{pi}^{-1}\sim10^3n^{1/2}~\text{s}$ and, moreover, it takes about a hundred $\omega_{pi}^{-1}$ seconds for an electrostatic shock [or $\omega_{pi}^{-1}(v_\text{shock}/c)$ seconds for a Weibel shock] to form and respond to the changing conditions; it takes even longer for particle Fermi acceleration. Thus, the synchrotron shock model should be used with great caution for $\Delta t$ as short as tens of milliseconds or less.

\subsubsection{Observables}

The above scalings can be used in equations for observables in Section \ref{s:obs} (note, most of the parameters turn out to be functions of the shock velocity alone). In particular, for $\Delta t\sim10^5\text{ s}$, i.e., about  $\sim1\text{ day}$ before a merger of an explosion, we have
\bea
\bar\gamma_e(\Delta t) &\simeq&  5.5\times10^{-2+2p_*}\ \chi_v^2\,\epsilon_e L_{s,39}^{2/5}t_{s,7}^{-9/5+p_*}n_{\text{ISM},0}^{-2/5}\Delta t_5^{-(p_*-1)}
\nonumber\\
&\simeq&  55\ \chi_v^2\,\epsilon_e L_{s,39}^{2/5}t_{s,7}^{-3/10}n_{\text{ISM},0}^{-2/5}\Delta t_5^{-1/2},\\
B(\Delta t) &\simeq& 2.1\times10^{-3+p_*}\ \chi_v\,\epsilon_B^{1/2} L_{s,39}^{1/5}t_{s,7}^{-(9-5p_*)/10}n_{\text{ISM},0}^{3/10} \Delta t_5^{-(p_*-1)/2}~\text{G}
\nonumber\\
&\simeq& 6.7\times10^{-2}\ \chi_v\,\epsilon_B^{1/2} L_{s,39}^{1/5}t_{s,7}^{-3/20}n_{\text{ISM},0}^{3/10} \Delta t_5^{-1/4}~\text{G},\\
\nu_s(\Delta t) &\simeq& 1.8\times10^{1+5p_*}\ \chi_v^5\,\epsilon_e^2\epsilon_B^{1/2} [(s-2)/(s-1)]^2 L_{s,39}t_{s,7}^{-(9-5p_*)/2}n_{\text{ISM},0}^{-1/2}\Delta t_5^{-5(p_*-1)/2}~\text{Hz}
\nonumber\\
&\simeq& 5.7\times10^{8}\ \chi_v^5\,\epsilon_e^2\epsilon_B^{1/2} [(s-2)/(s-1)]^2 L_{s,39}t_{s,7}^{-3/4}n_{\text{ISM},0}^{-1/2}\Delta t_5^{-5/4}~\text{Hz},\\
F_{\nu,\text{max}}(\Delta t) &\simeq& 8.9\times10^{1+p_*}\ \chi_R^3\chi_v\,D_{23}^{-2}\epsilon_B^{1/2} L_{s,39}^{4/5}t_{s,7}^{(9+5p_*)/10}n_{\text{ISM},0}^{7/10} \Delta t_5^{-(p_*-1)/2}~\text{Jy}
\nonumber\\
&\simeq& 2.8\times10^{3}\ \chi_R^3\chi_v\,D_{23}^{-2}\epsilon_B^{1/2} L_{s,39}^{4/5}t_{s,7}^{33/20}n_{\text{ISM},0}^{7/10} \Delta t_5^{-1/4}~\text{Jy},\\
F_{\nu}(\Delta t) &\propto& \nu^{-(s-1)/2}\Delta t^{-(5s-3)(p_*-1)/4} 
\nonumber\\
&\propto& \nu^{-(s-1)/2}\Delta t^{-(5s-3)/8}.
\label{fnu-app}
\eea
Here we remind that one should not use these scalings for too short $\Delta t$, as is discussed in the end of the previous subsection. 

One can reverse the argument here and ask: What physical parameters of the system can be inferred from observations? Obviously, if the spectral slope and the light-curve indexes of the flux
\beq
F_{\nu}(t)\propto \nu^{\beta_\nu}(\Delta t)^{\beta_t} 
\label{Fnu-obs}
\eeq
are measured in observations, one can readily determine the energy injection index $p_*$:
\beq
p_*=\frac{1-5\beta_\nu-2\beta_t}{1-5\beta_\nu},
\label{p-obs}
\eeq
where $\beta_\nu\not=1/5$, otherwise $s=3/5$ and $\beta_t=0$.

\subsection{Comparison with full numerical solution}
\label{s:ns}

\begin{figure}[b!]
\center
\includegraphics[angle = 0, width = 0.45\columnwidth]{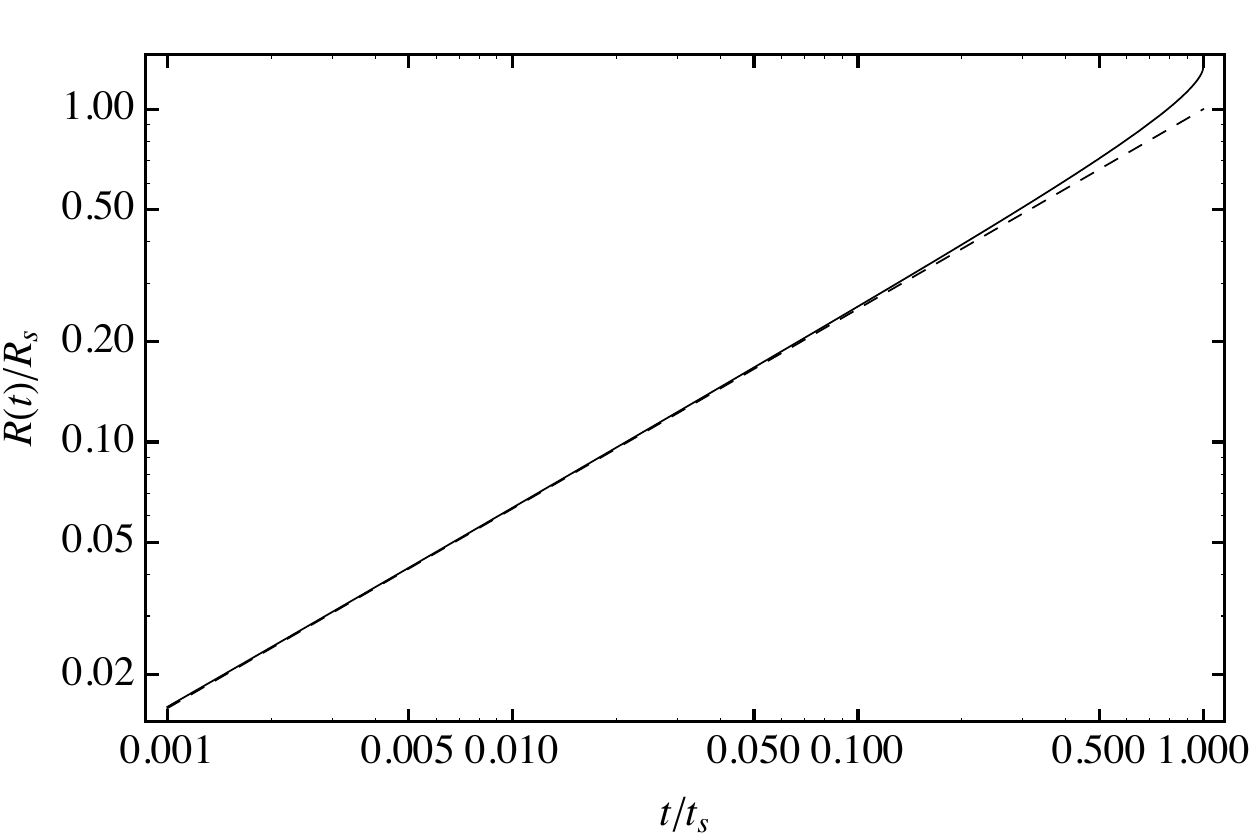}
\caption{Log-log plot of the shock radius driven by an expanding bubble as a function of time. The solid curve represents the full numerical solution of Eq. (\ref{master3}) and the dashed curve is the self-similar solution given by Eqs. (\ref{selfsim}), (\ref{ss-Rs}).
}
\label{numsolR}
\end{figure}

\begin{figure}[b!]
\center
\includegraphics[angle = 0, width = 0.45\columnwidth]{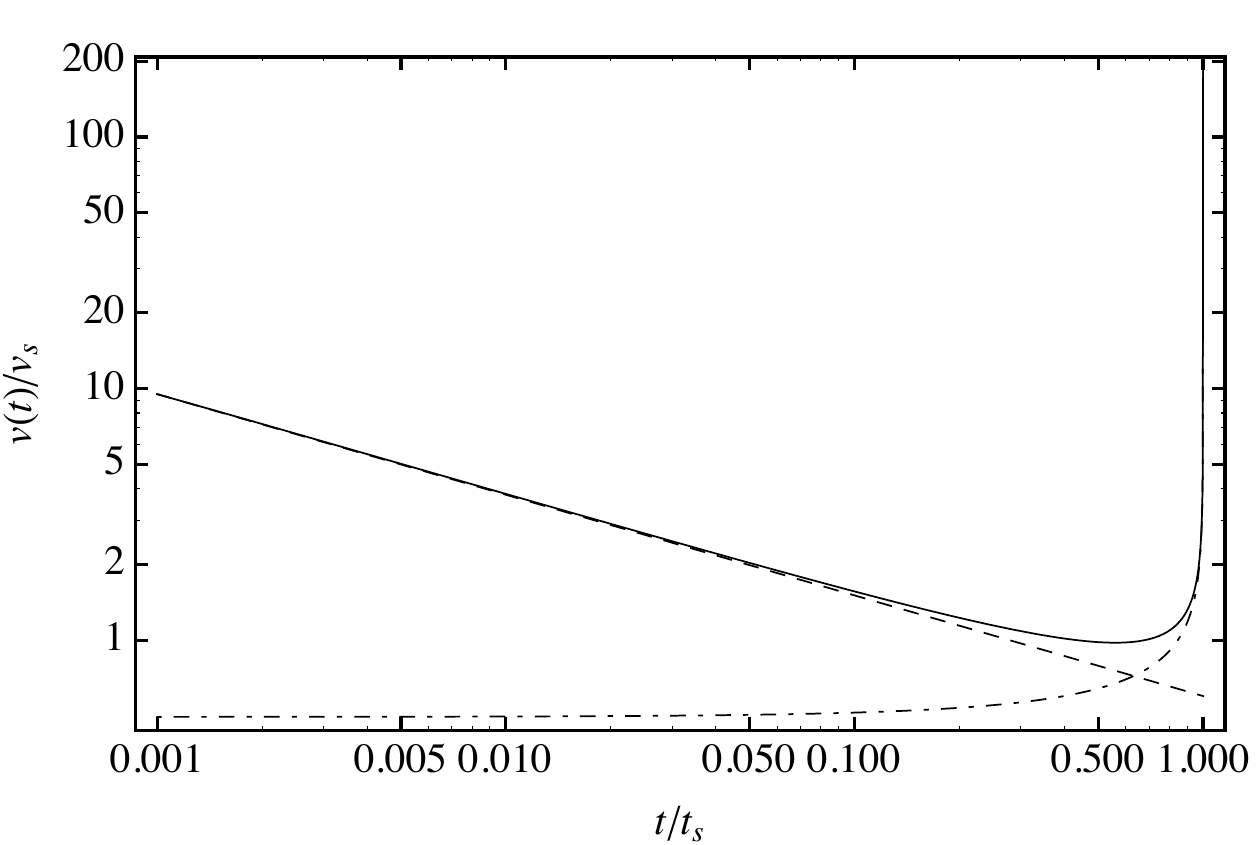}
\includegraphics[angle = 0, width = 0.45\columnwidth]{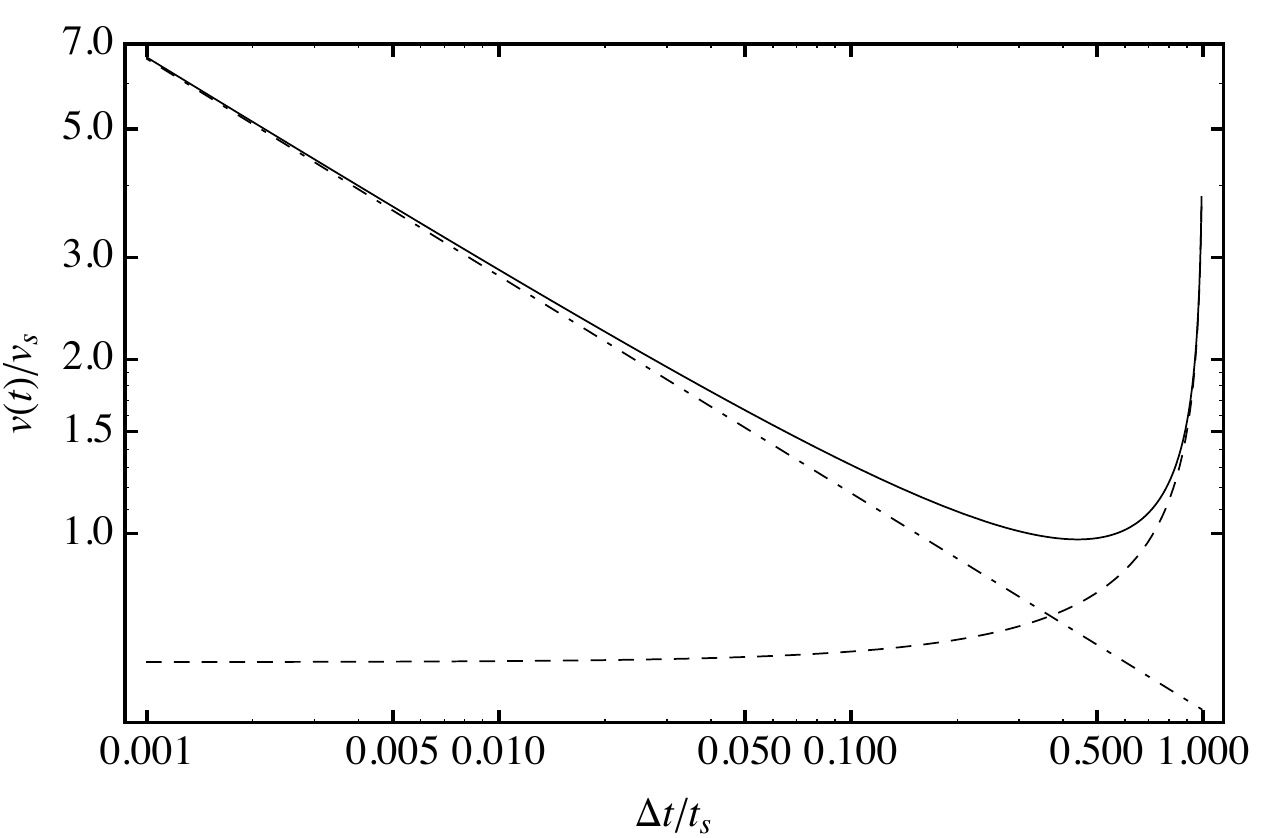}
\caption{Log-log plot of the driven-shock velocity as a function of time: $t/t_s$ (left panel) and $1-t/t_s$ (right panel); note that in the right panel, time evolution is from-right-to-left: the merger occurs at $\Delta t/t_s\to0$. The solid curve represents the full numerical solution of Eq. (\ref{master3}), the dashed curve is the self-similar solution given by Eqs. (\ref{selfsim}), (\ref{ss-Rs}) and $v_\text{shock}=\dot R(t)$, and the dot-dashed curve is the approximate finite-time-singular solution given by Eqs. (\ref{FTSfactors}), (\ref{vs}) with a correction factor included, see Eq. (\ref{chiV}).
}
\label{numsolV}
\end{figure}

The full numerical solution of equation (\ref{master3}) and the analytical solutions described in previous sections are shown in Figures \ref{numsolR} and \ref{numsolV}. Figure \ref{numsolR} shows the shock radius as a function of time. The solid curve is the exact numerical solution with $p=7/4$ and the dashed curve represents the approximate self-similar solution given by equations (\ref{selfsim}), (\ref{ss-Rs}) with index $p=0$, which corresponds to the early-time asymptotic of the realistic evolution (see discussion in Section \ref{s:ars}). The agreement of the self-similar solution with the exact one is remarkable. The noticeable deviation occurs only at the very late time, just before the coalescence time, but even then the difference is within a factor of order unity. Thus, the assumption of $R_\text{shock}\to const.$ as $t\to t_s$, used in Section \ref{s:ars} to derive the FTS solution, is justified. We determine numerically that the analytical solution underestimates the final radius (at $t=t_s$) of the shock by a factor of 
\beq
R_{s,\text{exact}}/R_{s,\text{selfsim}}\equiv\chi_R\sim1.33
\label{chiR}
\eeq
which we have included in the estimates of shock parameters and observables. 

Figure \ref{numsolV} shows the shock velocity as a function of time: $t/t_s$ (left panel) and of the time before singularity $\Delta t/t_s=(t_s-t)/t_s$ (right panel). The former elucidates the agreement of the exact numerical solution with the self-similar solution and the latter -- with the FTS analytical solution. Here the solid curve is the exact numerical solution, the dashed curve represents the approximate self-similar solution given by equations (\ref{selfsim}), (\ref{ss-Rs}) with index $p=0$ and the dot-dashed curve represents the approximate FTS solution. This analytical solution given by equations (\ref{FTSfactors}), (\ref{vs}) is found to overestimate the velocity by a factor of 2, i.e.,
\beq
v_{s,\text{exact}}/v_{s,\text{FTS}}\equiv\chi_v\sim0.51.
\label{chiV}
\eeq
Thus, the analytical solution represented in this figure is the one given by Eq. (\ref{vs}) with $v_s$ replaced with $\chi_v v_s$. Note the remarkable agreement between the exact numerical and analytical solutions.

\section{Conclusions}
\label{s:concl}

In this paper we considered the formation and evolution of an astrophysical bubble and the bubble-driven shock propagating in an ambient ISM of uniform density under the assumptions of spherical symmetry of the pressure balance throughout the system. The equation of the dynamics of the shock (and all other parameters) has been accurately derived for an arbitrary energy rate output by the central source, which we colloquially refer to as the ``luminosity law," $L(t)$. Furthermore, we derived the analytical solutions for two special cases: the self-similar scaling, $L(t)\propto t^p$, and the finite-time-singular case, $L(t)\propto(t_s-t)^{-p}$, where $t_s$ is the source life-time, and $p$ is a constant index. The latter ``explosive'' law can represent, for example, the energy output from a merging neutron star or magnetar binary. The analytical solution of the finite-time-singular type has not been derived before.

We have found that the dynamics of the bubble-driven shock is markedly different from the classical Sedov-von Neumann-Taylor solution even in the case of the ``explosive'' FTS behavior. The driven shock solutions only exist if the energy injection rate is not too low, namely the index $p$ shall exceed some critical value, $p>p_c$, which depends on the equations of state of the bubble and the ISM. In particular, for the standard ISM with the adiabatic index $\gamma=5/3$ and the bubble filled with the material with the relativistic equation of state (magnetized plasmas, electron-positron plasmas, electromagnetic radiation), $\gamma_b=4/3$, the critical value of $p_c$ is $-1.11$, for both types of solutions. Otherwise, if $p<p_c$, the bubble expansion is too slow to catch up with the outgoing shock. For $p>p_c$, the self-similar and FTS solutions for the shock radius and velocity are 
\beq
R_\text{selfsim}(t)\propto t^{(p+3)/5}, \quad v_\text{selfsim}(t)\propto t^{(p-2)/5}.
\eeq
\beq
R_\text{FTS}(t)\propto (t_s-t)^{(p-3)/5}, \quad v_\text{FTS}(t)\propto (t_s-t)^{(p-8)/5}.
\label{fts}
\eeq
This FTS solution is physical only if $p>3$; otherwise it is unphysical because it describes a converging shock. For $-1.11<p<3$ we have found a physically meaningful approximate solution 
\beq
R_\text{FTS}(t)\propto (t_s-t)^0, \quad v_\text{FTS}(t)\propto (t_s-t)^{-(p-1)/2}.
\label{fts-a}
\eeq

Interestingly, the derived solutions are also applicable to the class of systems with finite-time but non-singular luminosity laws: $L(t)\propto(t_s-t)^{-p}$ with $-1.11<p<0$. These are the systems which have a declining with time energy deposition rate and which are active for a finite time $t_s$. In the systems with $p>1$, the shock is accelerating as $t\to t_s$. Therefore, the bubble-shell interface may become Rayleigh-Taylor unstable if a less dense plasma of the bubble is pushing on denser ambient medium. Strong mixing is expected in this case. The overall dynamics may somewhat be affected, but the salient features should remain the same because the assumption of the pressure balance still holds. Another limitation of our analysis is the neglect of relativistic effects, which may be important if the shock velocity is singular at $t\to t_s$. We have also used the strong shock approximation throughout the analysis.

Finally, we made observational predictions. We calculated the emission light-curves of the bubble+shock systems for both self-similar and FTS cases. We predicted that one can deduce the luminosity law index $p$ from the spectral and temporal indexes, see Eqs. (\ref{Fnu-obs}), (\ref{p-obs}). Our results may be relevant to stellar systems with strong winds, merging neutron star/magnetar/black hole systems, as well as massive stars evolving to supernovae explosions.

\acknowledgements

One of the authors (MVM) is grateful to Sriharsha Pothapragada for useful suggestions. This work was
supported in part by DOE grant  DE-FG02-07ER54940 and NSF grant AST-1209665 (for MVM), and NSF grant AST-0907890 and NASA grants NNX08AL43G and NNA09DB30A (for AL).

\end{document}